\documentclass[iop]{emulateapj}

\usepackage{amssymb}                    % useful mathematical symbols 
\usepackage{graphics,graphicx}
\shorttitle{Study of the recurring dimming region detected at AR 11305}
\shortauthors{Krista \& Reinard}

\begin{document}


\title{Study of the recurring dimming region detected at AR 11305\\ using the Coronal Dimming Tracker (CoDiT)}


\author{Larisza D. Krista$^{1,2}$ and Alysha Reinard$^{1,2}$}
\affil{$^{1}$University of Colorado/Cooperative Institute for Research in Environmental Sciences, Boulder, CO 80205, USA}
\affil{$^{2}$National Oceanic and Atmospheric Administration/Space Weather Prediction Center, Boulder, CO 80305, USA}

\begin{abstract}

We present a new approach to coronal dimming detection using the COronal DImming Tracker tool (CODIT), which was found to be successful in locating and tracking multiple dimming regions. This tool, an extension of a previously developed coronal hole tracking software, allows us to study the properties and the spatial evolution of dimming regions at high temporal and spatial cadence from the time of their appearance to their disappearance. We use the Solar Dynamics Observatory/Atmospheric Imaging Assembly 193~\AA\ wavelength observations and Helioseismic and Magnetic Imager to study dimmings. As a demonstration of the detection technique we analyzed six recurrences of a dimming observed near AR 11305 between 2011 September 29 and October 2. The dimming repeatedly appeared and formed in a similar way, first expanding then shrinking and occasionally stabilizing in the same location until the next eruption. The dimming areas were studied in conjunction with the corresponding flare magnitudes and coronal mass ejection (CME) masses. These properties were found to follow a similar trend during the observation period, which is consistent with the idea that the magnitude of the eruption and the CME mass affect the relative sizes of the consecutive dimmings. We also present a hypothesis to explain the evolution of the recurrent single dimming through interchange reconnection. This process would accommodate the relocation of quasi-open magnetic field lines and hence allow the CME flux rope footpoint (the dimming) to expand into quiet Sun regions. By relating the properties of dimmings, flares, and CMEs we improve our understanding of the magnetic field reconfiguration caused by reconnection.
\end{abstract}


\section{Introduction}

Dimming regions are often referred to as transient coronal holes (CHs), due to their similar dark appearance in extreme ultraviolet (EUV) and X-ray wavelengths \citep{Hansen1974, Rust1983, Kahler2001}. Both dimmings and CHs are defined as reduced density regions of the lower corona dominated by ``open" magnetic flux \citep{Zarro1999, Thompson2000, deToma2005, Scholl2008}. However, the open flux is created through very different processes in these phenomena. While the creation of CHs is still debated (initial CH emergence near new active regions (ARs) and expansion via flux transport processes as suggested by \citet{Wang2010} or open field accumulation through interchange reconnection according to \citet{Fisk2005}), the creation of dimming regions is better understood: their appearance is linked to coronal mass ejection (CME) eruptions, during which magnetic field lines are dragged out by the ejected plasma into interplanetary space. This process creates a pseudo-open field through which plasma is evacuated and a transient CH (or dimming) is formed. 

It has been suggested that reduced emission in dimming regions could be partially caused by temperature variations \citep{Chertok2003, Robbrecht2010}; however, a study by \citet{Harrison2000} showed that the reduced intensities in dimming regions were observed at multiple wavelengths indicating mass evacuation. Furthermore, the Doppler observations in dimming regions shows plasma outflow which further proves that the main source of intensity reduction in dimmings is due to mass evacuation \citep{Harra2001, Tian2012}. Hence, dimmings become visible due to density and emission depletions near the AR as the plasma evacuates after the CME eruption. It has been suggested that dimmings are the footpoints of the erupting CME flux rope \citep{Thompson2000, Kahler2001}, which leads to the simultaneous appearance of two dimmings \citep{Thompson1998, Webb2000, Attrill2006, Gibson2008}. There have been numerous observations of double dimmings, but there are also a large number of dimmings with a more complex morphological configuration (one or several fragments of closely located dimmings). As these regions are associated with open magnetic field, dominant polarity is expected to be observed in both CHs and their transient counterparts. CHs are known to be dominated by a single polarity \citep{Wiegelmann2004, Scholl2008, Krista2009} and double dimmings also show clearly opposite dominant polarities in agreement with the CME flux rope footpoint association \citep{Webb2000, Attrill2006}. However, as we show in our work, singular complex dimmings may show a more mixed polarity, which could indicate several adjacent footpoints.

The strong link found between dimmings and CMEs \citep[e.g.,][]{Thompson1998, Webb2000, Thompson2000} has led to several studies aiming to estimate the amount of CME mass that corresponds to the evacuated plasma related to the dimming. \citet{Harrison2000} found that mass loss associated with dimmings could account for at least 70\% of the CME mass. \citet{Zhukov2004} found that approximately 50\% of the total CME mass in the low corona is contained outside of the dimming region. A recent study by \citet{Tian2012} showed that 20\%-60\% of the CME mass is associated with the dimming. It is important to note that CME mass calculations bear large errors due to the uncertainties associated with the emission measure distribution and the angle at which the CME is observed, while dimming area measurements are largely dependent on the detection method. Hence, there is still a need for improvement in these measurement efforts. Another study by \citet{Reinard2009} found that the dimming-CME relationship can also hold some predictive power: it was found that dimming-associated CMEs tend to have a large range of speeds (with a 50\% chance of a fast CME), while CMEs with no associated dimmings do not exceed 800 km s$^{-1}$. Also, dimming-associated events were found more likely to be accompanied by flares with large magnitudes.

Several methods of dimming detection use base-difference images to determine the location and size of dimmings \citep[e.g.,][and references therein]{Reinard2008, Attrill2010}. This approach shows the intensity changes relative to the pre-event image chosen by the observer. While this method is highly successful in locating dimmings it also carries uncertainties, as it is based on temporal changes in the intensity, which could be caused by plasma ``displacement" (i.e., by moving features) as well as ``removal" (i.e., evacuation of matter). Furthermore, as the eruption progresses, the general drop in the intensity in the areas surrounding the AR leads to the difference images often showing large dimming areas encompassing the whole AR. These boundaries are not in agreement with the clearly defined boundaries seen on the original (non-difference) EUV images. For these reasons we use the original calibrated EUV images to determine the dimming areas using the Coronal Dimming Tracker (CODIT) algorithm. CODIT was developed from the Coronal Hole Evolution (CHEVOL; \citeauthor{Krista2011} \citeyear{Krista2011}) algorithm, which was designed to detect and track the morphological evolution of emerging CHs. Since dimming regions are essentially transient CHs, the method was straightforward to adapt. In the work presented here we use the high spatial and temporal resolution data available from the Solar Dynamics Observatory/Atmospheric Imaging Assembly (AIA; \citeauthor{Lemen2012} \citeyear{Lemen2012}) and Helioseismic and Magnetic Imager (HMI; \citeauthor{Scherrer2012} \citeyear{Scherrer2012}). The details of the algorithm are discussed in Section \ref{method}. 

We have analyzed several dimming regions using CODIT, but in this study we concentrate on a particularly interesting case of a recurring dimming observed at AR 11305 between 2011 September 29 - October 2. Our goal is to investigate the relationship between the erupting flares, the associated dimming regions that appeared immediately after the flares, and the CMEs that were detected off-disk in coronagraph images (Section \ref{results}). The present study is based on a single AR and its recurrent eruptions, which gives us information on the energy stored in the AR and how it is released over time through triggering effects. The recurrence of the single dimming region also raises a question about the processes that lead to the repeated opening and closing of the magnetic field near AR 11305. We repeatedly observed a single rather than a double dimming, which makes the CME flux-rope model interpretation more difficult. Previous hypotheses indicate that interchange reconnection may play an important role in the evolution of open magnetic field region boundaries \citep{Fisk2005, Krista2011} and the creation of dimming regions \citep{Attrill2006, Gibson2008}. Using the interchange reconnection approach we present a hypothesis to explain the repeated emergence of a dimming near AR 11305 (Section \ref{conclusions}).

\section{Observations and Data Analysis}
\label{method}

\subsection{Observations}

Dimming regions are best observed in EUV or X-ray images. The additional use of magnetograms are also important as they provide us with information on the photospheric magnetic field in the dimming regions, which are thought to be the CME flux-rope footpoints. We developed and tested the CODIT algorithm using SOHO {\it Extreme Ultraviolet Imaging Telescope} (EIT; \citeauthor{Delaboudiniere1995} \citeyear{Delaboudiniere1995}) 195~\AA\ images and Michelson Doppler Imager (MDI; \citeauthor{Scherrer1995} \citeyear{Scherrer1995}) magnetograms. SOHO/EIT data are available from 1996 until today; however, the temporal cadence was greatly reduced and MDI magnetograms observations were discontinued after April 2011. For this reason and the availability of better quality data from the SDO we use the latter for dimming observations after 12 May 2010 (from which date both SDO EUV data and magnetograms are available in the JSOC data center). We have adapted the algorithm to work with SDO/AIA 193~\AA\ observations together with the HMI magnetograms. The SDO observations provide unprecedented high spatial and temporal cadence observations which allow us to study dimmings in more details. In our work we use the HMI line-of-sight (LOS) magnetograms, which have a spatial resolution of 1 arcsec/pixel, a time cadence of 45 seconds and a precision of 10 G in a dynamic range of $\pm$4 kG. The AIA ensemble observes the Sun and its environment to 1.3 solar diameters in multiple wavelengths, at a 1.5 arcsec/pixel resolution and 12 second cadence. We use the 193~\AA\ observations which show the dimming boundaries at the highest clarity compared to other wavelengths. The evolution of the dimming regions is not rapid enough to require a 12 second cadence, hence to decrease computational time we use a cadence of $\sim$7 minutes where data availability allows.

In order to detect dimming-related CMEs we use the STEREO/SECCHI COR 2 coronagraph observations \citep{Howard2008}. COR2 is an external-occulter Lyot coronagraph that observes the weak coronal signal in visible light from 2 to 15 solar radii. The spatial resolution of the instrument is 14.7 arcsec/pixel and the temporal cadence is 15 minutes. We used the STEREO-A or STEREO-B COR2 instrument data depending on when the angle between the satellite and AR location was closest to 90$^\circ$. This angle was within $\sim10^\circ$ for each CME observation.

\subsection{Dimming Detection}

\subsubsection{Previous Methods}

\citet{Alipour2012} describe the automated detection of small-scale EUV dimmings related to micro flares using a feature-based classifier. The method is based on STEREO/EUVI and SDO/AIA 171 \AA\ images and uses a learning system based on statistical learning theory. The dimming intensity threshold is chosen as the full width at half-maximum of the intensity histogram for each space-time slice analysed. While this method is promising, it is not yet adapted for the detection of larger dimmings. Other methods use running or base-difference images. In a sequence of observations, running-difference images are created by subtracting the previous frame from a given frame. In the case of base-difference images, a pre-event reference image is subtracted from all the frames in the sequence. \citet{Reinard2008} used base-difference images to detect dimming regions in SOHO/EIT 195~\AA\ images. The method requires the user to define a region of interest (ROI), where the dimmings are identified as pixels with intensities lower than $1\sigma$ below the mean intensity of the whole difference image. This method was successful in detecting 96 dimmings between 1998-2000. Another approach is the Novel EIT wave Machine Observing (NEMO) algorithm \citep{Podladchikova2005}, which detects the occurrence of coronal waves and dimmings by finding a significant perturbation in the statistical distribution of the pixels in running-difference images. Once the dimming occurrence is identified, the dimming is extracted using base-difference rather than running-difference images, which allows the study of the dimming evolution rather than the variation in the dimming. The method involves the selection of the 1\% darkest pixels from the base-difference image, then region-growing is applied to determine the full extent of the dimming area. The region-growing is done to the extent where the dimming is still simply-connected and is within a region defined by a maximal intensity threshold. This method is most likely to have difficulty detecting dimming events that are not accompanied by EUV waves or have a higher noise level, since the dimming signature might not be significant enough in the pixel distribution. Also, using this method dimmings and EUV waves could be a challenge to distinguish. \citet{Attrill2010} further developed the NEMO method by changing the initial detection step, where instead of using the running-difference images, they use the statistical properties of the original image pixels. This step was introduced to reduce computational time. In order to improve the detection of smaller dimmings they also used sub-images to increase pixel statistics. The dimming extraction step was also modified from NEMO: the mean value of the pre-event base-difference reference image is calculated and the $1\sigma$ threshold is used to determine the dimming pixels.

While these methods have all been successful in detecting dimming regions, both running and base-difference images introduce methodological artefacts. Moving EUV wave fronts in running-difference images can be mistaken for dimmings as a dark region will appear in the location where the wave front was in the previous frame. The decreasing intensity in the flare location will also lead to a dark feature and can be falsely included as part of the dimming. A slight increase in the dimming intensity, however, will lead to a bright region in the running-difference image and can falsely modify the dimming boundary and size. If the dimming intensity does not change in consecutive frames, the dimming will be completely absent in the difference image. Base difference images can also introduce artefacts - the resulting dimming regions are dependent on the user-chosen base image that shows how the dimming changed relative to one pre-event image over time. The overall change of intensity in the AR also influences the extent of the detected dimming, producing considerably larger dimmings than those seen in EUV images. Solar rotation also has to be corrected for in base-difference imaging in order to avoid bright/dark eastern edges appearing due to rotational displacement of dark/bright features \citep{Chertok2003}. Plasma movement (e.g., loop displacement) can also be mistaken for dimming regions using these methods. Furthermore, none of the previously discussed methods use magnetograms which could provide information on the photospheric magnetic field distribution in the CME footpoints. For these reasons we developed a method using the original, non-difference EUV images and the corresponding magnetograms for the comprehensive study of dimming regions.

\subsubsection{Coronal Dimming Tracker (CODIT)}

The CODIT algorithm is based on the Coronal Hole Evolution (CHEVOL; \citeauthor{Krista2011} \citeyear{Krista2011}) algorithm, which was designed to detect and track the morphological evolution of emerging CHs. Since dimming regions are essentially transient CHs, the method was straightforward to adapt. However, the rapid changes in morphology and the sometimes higher intensity of dimmings compared to CHs also had to be taken into consideration to best suit dimming detection. 

CODIT can be used at different wavelengths and with different instruments, however the Fe~{\sc xii} wavelength (e.g., 193 and 195~\AA) was found to be most suitable due to the high contrast between dimmings and the quiet Sun (QS). In the present work we use the SDO/AIA 193~\AA\ observations and the HMI magnetograms. We use a time cadence of 7 minutes, which is adequate to follow the morphological evolution of dimmings and requires a relatively short computational time (e.g., 3 hr processing time for a 13 hr observation period). Before the dimming detection is performed, the AIA data are calibrated using the {\it aia\_prep} routine (available in the SolarSoft software package; \citeauthor{Freeland1998}  \citeyear{Freeland1998}), then the EUV disk image is transformed to a Lambert cylindrical equal-area projection map. The Lambert maps aid the thresholding process and the projection also corrects for LOS distortion of the regions on the spherical disk image, hence the correction also aids the study of the real boundary evolution of dimming regions. It is important to note that the Lambert projection is most accurate within $\pm$60$^\circ$ latitude and longitude from the disk center, and consequently the dimming boundaries are most reliable within these limits. 
\begin{figure*}[!t]
\centerline{
\includegraphics[scale=0.4]{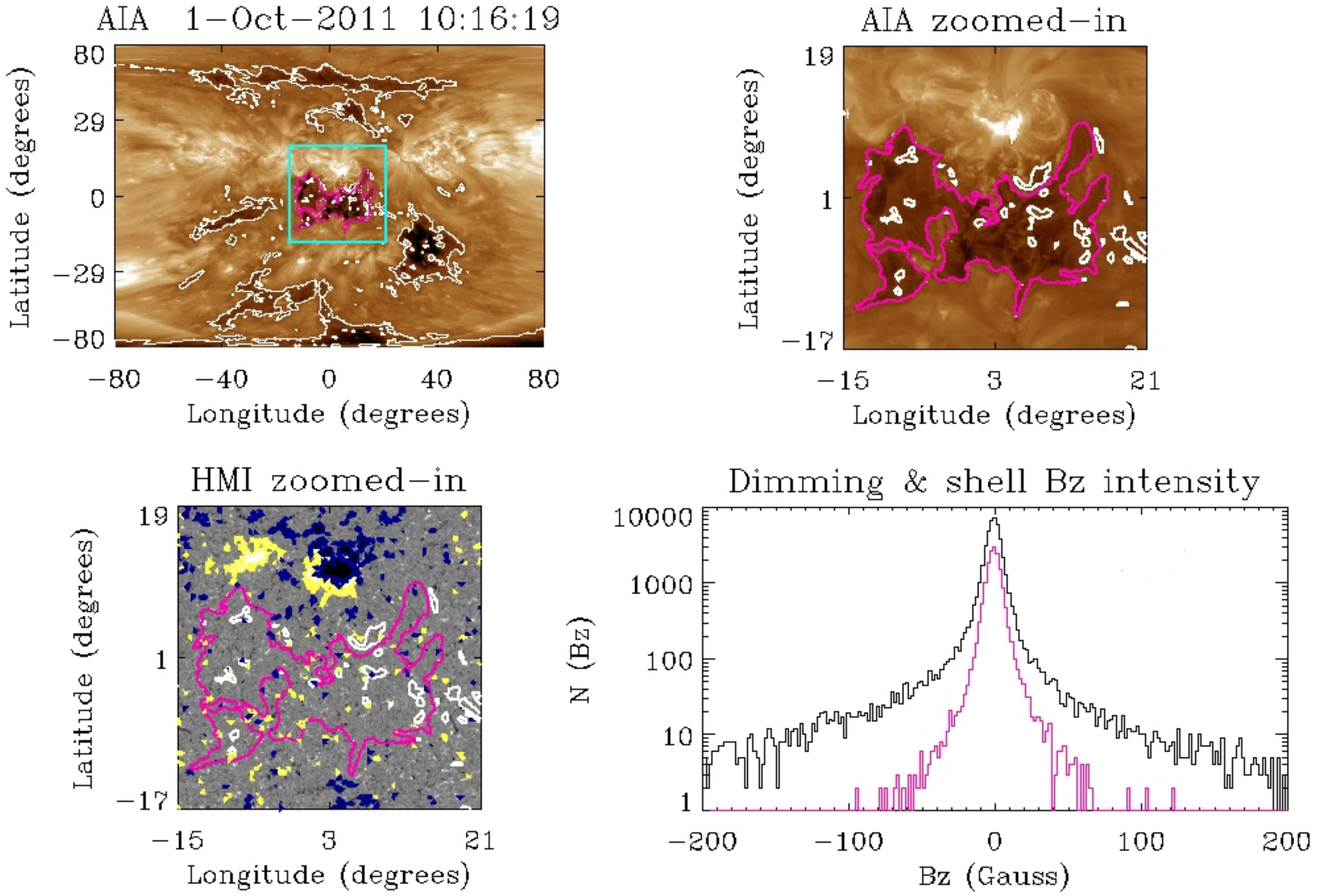}}
\caption{Example of CODIT detection products. Top left panel: the AIA 193~\AA\ Lambert map of 1 October 2011 10:16 UT, with the magenta contours highlighting the dimming region within the region of interest (ROI; green box), the white contours show all the regions with the same intensity threshold. Top right panel: a close up of the AIA ROI with the dimming contours highlighted. Bottom left panel: ROI in the corresponding HMI magnetogram, with the same dimming contours overlaid. The yellow and blue contours correspond to magnetic field strength above 50 G and below -50 G, respectively. Bottom right panel: histogram showing the magnetic field distribution inside the dimming region (magenta curve) and the surrounding regions within the ROI, denoted by ``shell" (black curve).}
\label{dim1}
\end{figure*}  
The dimming detection method is based on a local intensity thresholding approach which differentiates low intensity regions (e.g., dimmings or CHs) from the higher intensity QS. During the thresholding process different size windows are stepped through the Lambert image to extract the local intensity distribution where the local threshold is automatically identified. The most frequent threshold found in all the sub-images is then used as an intensity threshold for the whole Lambert map. The thresholding method is described in detail in \citet{Krista2009} for the detection of CHs. The small-scale CH evolution tracking algorithm (CHEVOL), on which CODIT is based is described in detail in \citet{Krista2011}. CODIT is a semi-automated detection method, which allows the user the define the size and location of the ROI (see light-green box around the dimming in Figure \ref{dim1}) where the dimming is detected. The user also provides the time at which the dimming was first observed, the observation period and the image analysis frequency. Once these inputs are provided the algorithm automatically detects the dimming and follows it across the solar disk over the observation period. On some occasions pre-existing dark regions (other dimmings or CHs) can be observed on the solar disk at the time when the dimming of interest appears. Using our method, the unrelated dark areas are excluded by the ROI. However, it needs to be noted that in the case of recurrent dimmings (like those discussed in this paper), if a dimming appears on the ``ashes'' of a previous dimming, our tool does not differentiate between the pre-existing dimming and the one appearing newly in the same location (running or base-difference images would exclude the pre-existing dimming). Our tool tracks the new dimming as it grows out of the pre-existing dimming.

An example of a dimming detection is shown in Figure \ref{dim1}: the upper left panel shows the AIA 193~\AA\ Lambert map, with the magenta contours highlighting the dimming within the ROI. The upper right panel shows a close-up of the ROI. Here, the AR with bright loop structures can be seen north of the dimming, and the flare site can be clearly seen as the saturated, small white region. The ROI is shown in the corresponding HMI magnetogram in the lower left panel, with the dimming contours overlaid. The yellow and blue regions highlight magnetic field strength above 50 G and below --50 G, respectively. The polarity distribution of the AR can also be seen as concentrated negative magnetic fields (blue) in the northern parts, and mainly positive magnetic fields (yellow) in the southern parts of the AR. The lower right panel shows the magnetic field distribution inside the dimming region (magenta curve) and the surrounding regions within the ROI (black curve). Studying the magnetic field distribution within the dimming region is particularly important in order to determine which of the CME flux rope footpoints are observed and how they change over time. The photospheric magnetic field can also indicate the complexity of the CME magnetic field configuration. For example a single dimming with mixed photospheric polarity (e.g., the dimming shown in Figure \ref{dim1}) suggests a configuration that could be more complex than the two-footpoint flux rope configuration associated with a double-dimming of opposite polarity.

Dimmings evolve faster and often in a more complex fashion than CHs, therefore a ``memory feature" has been added to CODIT in order to track the complex changes in the dimming morphology. This process compares the dimming region in one observation to the next and determines if the dimming merged with another region or split into smaller satellite dimmings. This helps to track the changes in the dimming morphology without losing information on the disconnecting dimming fragments. 

\section{Results}
\label{results}

In this study we have concentrated on AR 11305, which produced several flares, dimmings and CMEs during our observation period. The most intriguing aspect of the AR was the recurrence and increasing size of the dimmings. Due to the AR location near the equator and the flare activity during its passage across the central meridian, we had a prime observational opportunity with minimal LOS effects. The location of the STEREO A and B satellites was also at an optimal angle allowing a close to 90$^\circ$ view of the corresponding CMEs in the coronagraph images, which minimizes the CME mass calculation errors due to plane of sky assumption \citep{Vourlidas2000, Colaninno2009}.

\subsection{Observations of Dimming Regions}

We detected six dimming regions at AR 11305 between 29 September - 2 October 2011. In the movies available online (e.g., dim1.mp4) we show the evolution of each dimming. We can see that the first dimming (dim1.mp4) was the smallest of the six dimmings, and it shrank and disappeared before the second dimming appeared in the same location (dim2.mp4). The second dimming shrank as well, but did not disappear before the third dimming appeared and assumed a very similar morphology (dim3.mp4). There is little left of the third dimming by the time the fourth dimming appears (dim4.mp4), which was the largest in its area in comparison to all six dimmings. The fourth dimming began to shrink in size, but was still quite prominent when the fifth dimming appeared (dim5.mp4). The fifth dimming was the second largest in its area and appeared to have a similar morphological evolution to the fourth dimming. While all dimmings appeared to spread in a southward direction from the AR, the fourth and fifth dimmings spread in an eastward direction as well and shrink back in a westward direction. The fifth dimming shrank and practically disappeared by the time the sixth dimming occurred (dim6.mp4). The sixth dimming also showed some eastward directed growth, however it mostly shrank northward. The dimmings, as they reduced in size, did on some occasions separate into two regions, but they still showed a mostly mixed polarity. While all dimmings appeared in a similar location, their morphological evolution and largely mixed polarities indicate the complex evolution of the magnetic field which is not typical of the simple CME flux rope model that leads to two well separated footpoints with opposite polarities.

\subsection{Dimming Regions and the Corresponding Flares and CMEs}

\begin{figure}[]
\centerline{
\includegraphics[scale=0.68,  trim=180 100 200 90, clip]{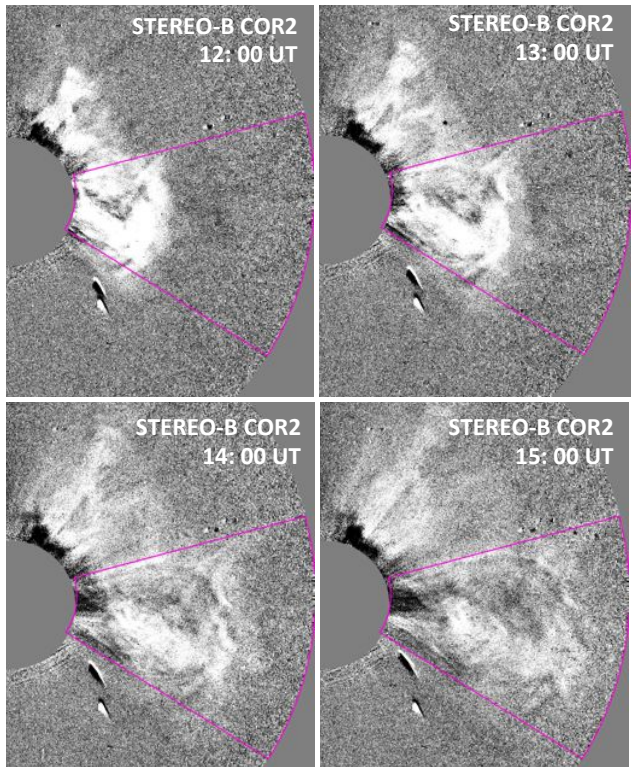}}
\caption{The CME related to the 9:59 UT 1 October 2011 flare, detected at four consecutive times: 12:00 and 13:00 UT (top left and right), 14:00 and 15:00 UT (bottom left and right). The magenta sectors were selected as the CME outline by the author.}
\label{cme1}
\end{figure}

In Table 1 the flare magnitudes are listed for each consecutive eruption (the flare magnitudes listed here were taken from the flare catalogues of the NOAA GOES satellite and the Hinode satellite). Flare classification is based on the X-ray radiation intensity. In the case of the observed flares an increasing maximum flare intensity was observed with each consecutive eruption. However the last (sixth) flare was smaller and perhaps indicated a reduction in the energy output. The maximum dimming areas show a similar increasing trend with each eruption. However, the largest dimming was the fourth; the fifth dimming was similarly very large at its peak area, and the sixth dimming was considerably smaller. Associated CMEs were only detected for the last four dimmings. The calculated CME masses were found to follow a similar trend as the maximum dimming areas: the CME mass increased from the third to the fourth dimming, then decreased to the fifth with a further decrease to the sixth. The similar trend in the magnitude of dimming areas and CME masses supports previous works \citep[e.g.,][]{Harrison2000, Zhukov2004, Tian2012} that suggest CME masses originate partly from dimming regions.
\\
\begin{table}[t]
\centering
\caption{Flare, Dimming, and CME Properties Related to the Six Dimming Recurrences at AR 11305}
\begin{tabular}{ c c c c c }
\hline \hline
\#  & Flare time & Flare & Max($A_{dim}$) & $m_{CME}$ \\
  & [UT] & magnitude  &  [$10^{4} Mm^{2}$] &  [$10^{15} g$] \\ \hline
1 & 18:09 Sep 29 & C5.6 & 0.7  & --- \\
2 & 04:00 Sep 30 & C7.7 & 1.5  & --- \\
3 & 19:06 Sep 30 & M1 & 1.7  & 1.37 \\
4 & 09.59 Oct 1 & M1.2 & 5.3  & 3.43 \\
5 & 00:50 Oct 2 & M3.9 & 4.8  & 1.42 \\
6 & 21:49 Oct 2 & C7.6 & 3.1  & 1.29 \\
\end{tabular}
\end{table}
%Flare infos from: http://msslxr.mssl.ucl.ac.uk:8080/SolarB/eisflare2011.jsp      ->Hinode EIS flares
%Sept 29 flare info from: http://st4a.stelab.nagoya-u.ac.jp/hinode_flare/hinode_html/201109_event.html      ->Hinode XRT flare
The CME masses were calculated using the algorithm developed by \citet{Colaninno2009} based on the method of \citet{Billings1966}. The algorithms used are the {\it scc\_calc\_cme\_mass} and {\it scc\_cme\_massimg2total} procedures, which are available through SolarSoft. 
%at http://secchi.nrl.navy.mil/wiki/pmwiki.php?n=Main.DataProcessingAndAnalysis. A tutorial can be found at http://sohowww.nascom.nasa.gov/solarsoft/stereo/secchi/doc/cme\_mass\_tutorial.html. 
The STEREO/COR2 images are first calibrated using the {\it secchi\_prep} routine (available in SolarSoft) and converted to units of mean solar brightness. Then a pre-event image is chosen to be the base image which is then subtracted from the image where the CME is visible. This step removes the background F-corona, static K-corona and residual stray light, and reveals the brightness change caused by the CME that has appeared in the corona. The inclination of the CME is not known precisely and hence it is assumed that all the electrons are located along the plane of sky. The number of electrons is then estimated as the ratio of the observed brightness and the estimated brightness of a single electron at a given angular distance. For more details see Section 2 of \citet{Colaninno2009}. An example CME detection is shown in Figure \ref{cme1}. This CME, observed by STEREO-B, corresponds to the M1.2 flare that occurred at 9:59 UT 1 October 2011 and the dimming shown in Figure \ref{dim1}. The CME was detected at four consecutive times, showing its expansion (top left to right: 12:00, 13:00 UT, bottom left to right: 14:00, 15:00 UT, 1 October 2011). The CME location is highlighted with the magenta sectors shown in the figure (the location and angular extent of the sectors was chosen by the author). The outer edge of the sector was set to the maximum field of view to make sure that the faint parts of the CME front were included in the mass calculation too. The masses shown in Table 1 correspond to the mass of each CME at the time when its front had passed 6 R$_{Sun}$. At this stage the CMEs are developed enough to show their full structure but are still within the COR2 field of view, allowing the mass of the full CME to be calculated.

\section{Discussion and Conclusions}
\label{conclusions} 

\begin{figure}[!t]
\centerline{
\includegraphics[scale=0.45, trim=100 0 100 0, clip]{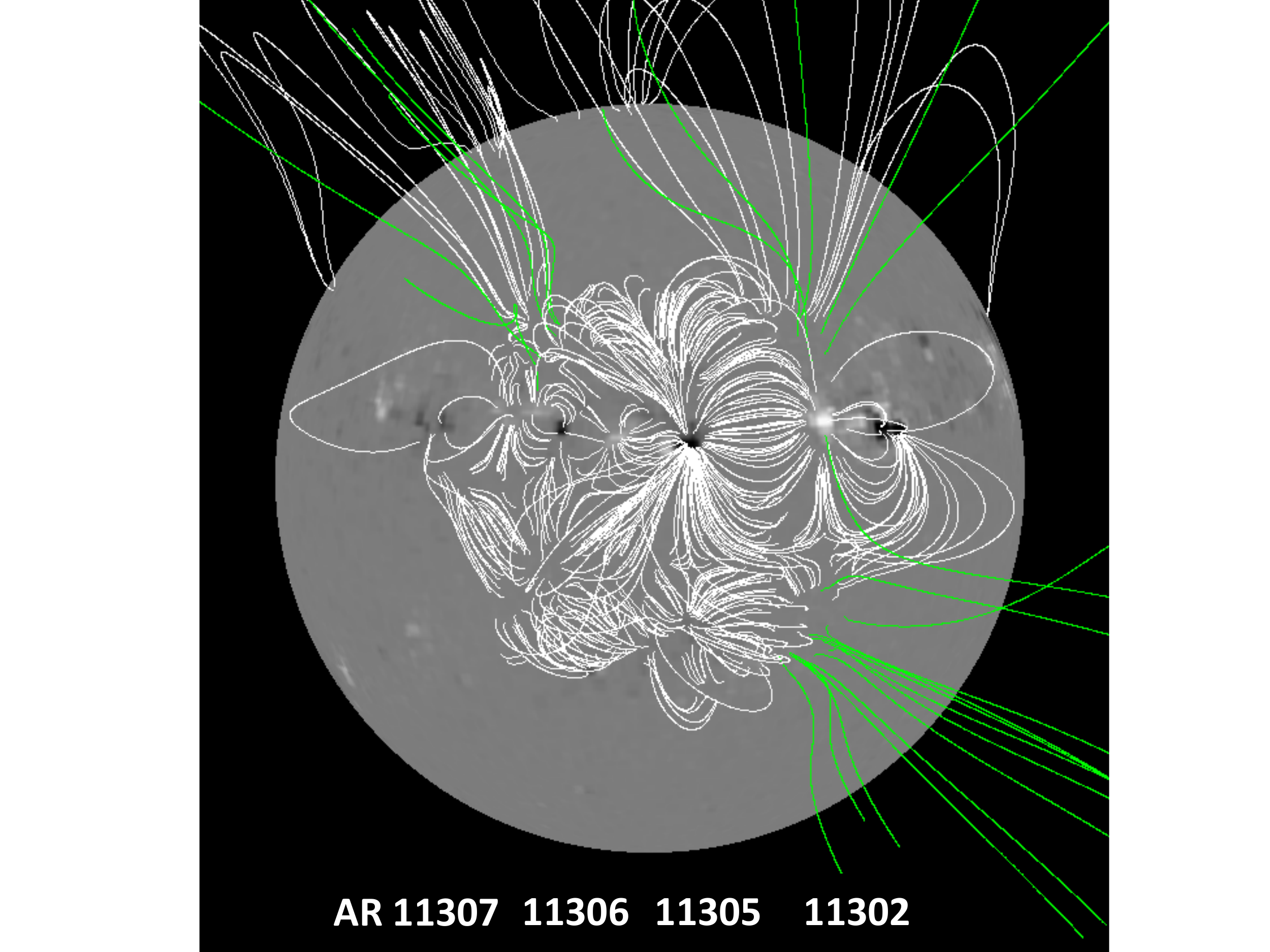}}
\caption{The PFSS plot showing the magnetic field lines connecting AR 11305 (disk center) to nearby quiet and active regions on 1 October 2011. The white lines show field lines that connect back to the solar surface, and the green lines show open magnetic field lines. The ARs are listed underneath their location in the image.}
\label{pfss}
\end{figure}

Figure \ref{pfss} shows the large scale configuration of the magnetic field for AR 11305 and the surrounding regions. We used the algorithm developed by M. L. DeRosa (description and algorithms available at {\it http://www.lmsal.com/$\sim$derosa/pfsspack/} and through SolarSoft) which uses the potential field source surface (PFSS) model to calculate the trajectories of magnetic field lines from a Carrington rotation magnetogram. The figure is centered on AR 11305 at 6:04 UT 1 October 2011. Black and white regions on the solar disk correspond to negative and positive polarity magnetic fields. The white lines correspond to closed, and the green lines to open magnetic field lines. It is clear from the image that AR 11305 is connected to surrounding regions all around, including nearby ARs (on the east: AR 11306, on the west: AR 11302). AR 11305 erupted six times, which indicates that there was built-up energy that got released gradually with each eruption. Eruptions in AR 11305 were mostly triggered by eruptions in nearby regions. Similar occurrences of sympathetic flaring have been previously discussed extensively by several authors \citep[e.g.,][]{Torok2011,Jiang2011,Schrijver2011,Moon2002}. An example of such sympathetic flaring occurred when a flare erupting in AR 11302 (west of AR 11305) at $\sim$ 9:04 UT 1 October 2012 appears to have destabilized the magnetically connected AR 11305 and given rise to the flare occurring at $\sim$ 9:59 UT the same day.

\begin{figure}[!t]
\centerline{
\includegraphics[scale=0.47, trim=150 50 150 50, clip]{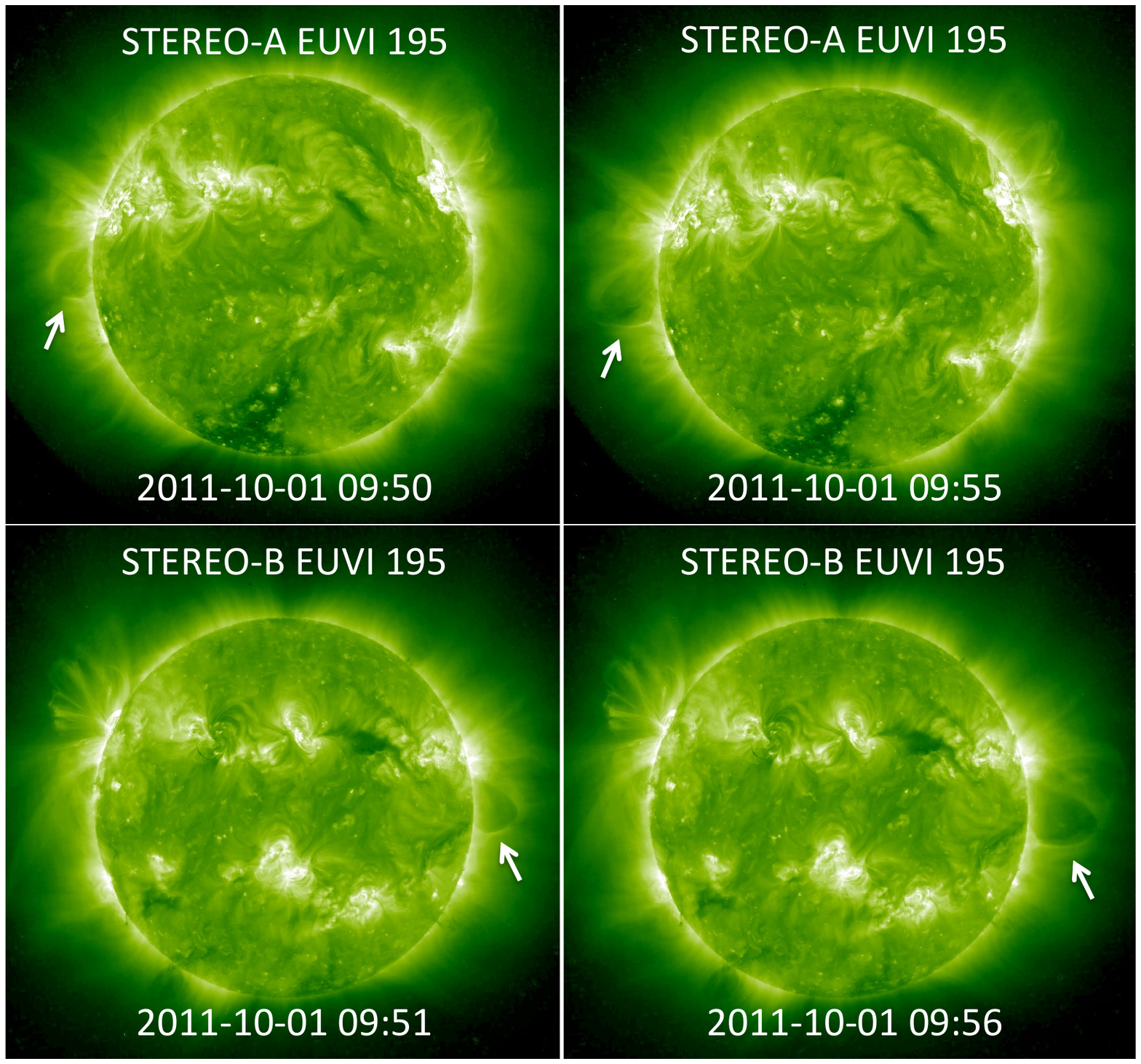}}
\caption{Top left and right: STEREO-A EUVI images showing the location of the southward deflecting CME loop structure at 9:50 and 9:55 UT 1 October 2011, respectively. Bottom left and right: STEREO-B EUVI images showing the location of the CME loop structure at 9:51 and 9:56 UT 1 October 2011, respectively. The arrows point to the loop structure that expands and becomes a CME. This CME corresponds to the dimming shown in Figure \ref{dim1} and the CME shown in Figure \ref{cme1}. The reader is encouraged to view the corresponding EUVI movies: EUVI-A\_195.mp4 and EUVI-B\_195.mp4.}
\label{euvi}
\end{figure}

Magnetic field lines can also be seen extending southward from AR 11305. A closer look at the magnetic polarity distribution in AR 11305 (see HMI sub-image in Figure \ref{dim1}, north of the dimming outline) shows the AR dipole structure: strong negative fields in the northern part of the AR and strong positive fields in the southern part. We suggest that the repeated dimming appearance south of the AR 11305 might be caused by the positive polarity footpoints of the AR loop structure expanding and extending southward into the mixed polarity QS. Interchange reconnection between the southern strands of the AR loops and the randomly oriented small dipole structures in the QS could accommodate the ``displacement" of the loop footpoints, consequently the CME flux-rope footpoints spread out in a southward direction. As the CME eruption drags the field lines out into the heliosphere these displaced footpoints become the dimming region visible as dark patches in the EUV images. Also, the relationship between the CME mass and the dimming size (as seen in Table 1) suggests that more massive (i.e., larger) CMEs cause more interchange reconnection and thus lead to larger dimmings. We suggest that the dimmings spread in a southward direction as there is mixed-polarity QS south of the AR. The negative polarity magnetic fields are concentrated in the northern parts of the AR and the regions north of the AR are also dominated by the same (negative) polarity, which inhibits the movement of the CME loop footpoints in a northward direction. Consequently the dimming cannot spread northward. Inspecting STEREO/EUVI 195~\AA\ images we found that all CMEs were deflected in a southward direction which is in agreement with the southward forcing of the dimming expansion. Figure \ref{euvi} shows this southward directed deflection of the CME loop structure in the EUVI observation taken by STEREO A and B between 9:50 - 9:56 1 October 2011 (see movies EUVI-A\_195.mp4 and EUVI-B\_195.mp4). The CME showed here corresponds to the dimming shown in Figure \ref{dim1} and the CME shown in Figure \ref{cme1}.

\begin{figure}[!th]
\centerline{
\includegraphics[scale=0.29, trim=0 100 0 220, clip]{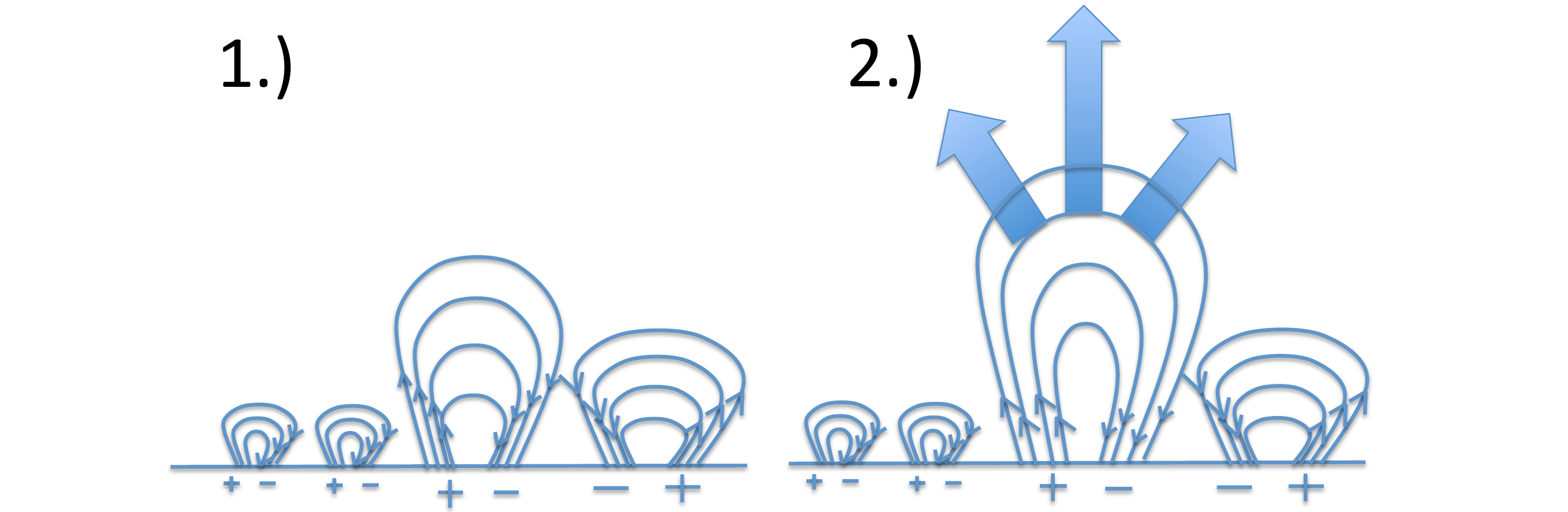}}
\centerline{\hspace*{0.015\textwidth}
\includegraphics[scale=0.29, trim=0 30 0 10, clip]{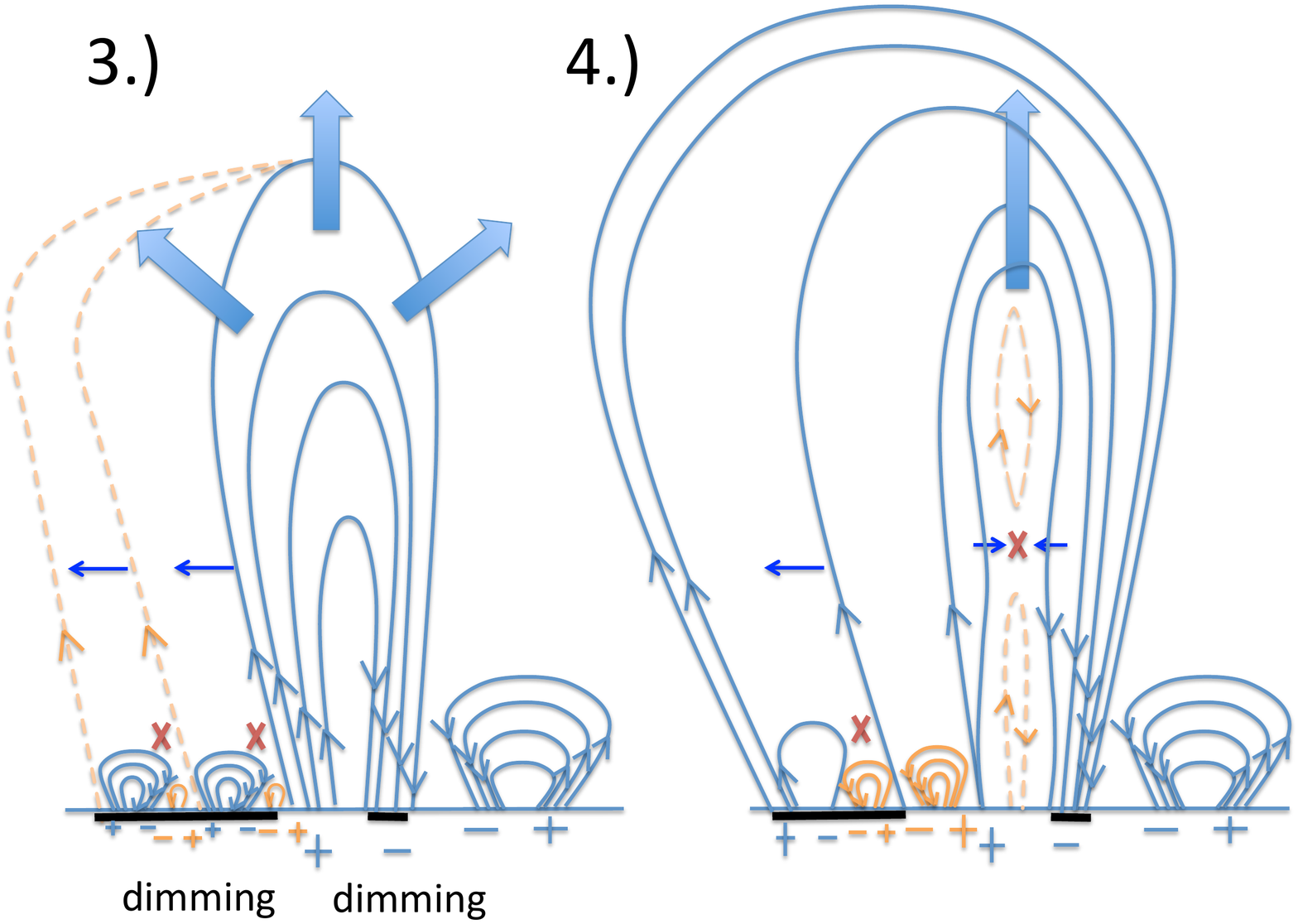}}
%\vspace{0.1}
\centerline{\hspace*{0.015\textwidth}
\includegraphics[scale=0.26, trim=105 30 270 30, clip]{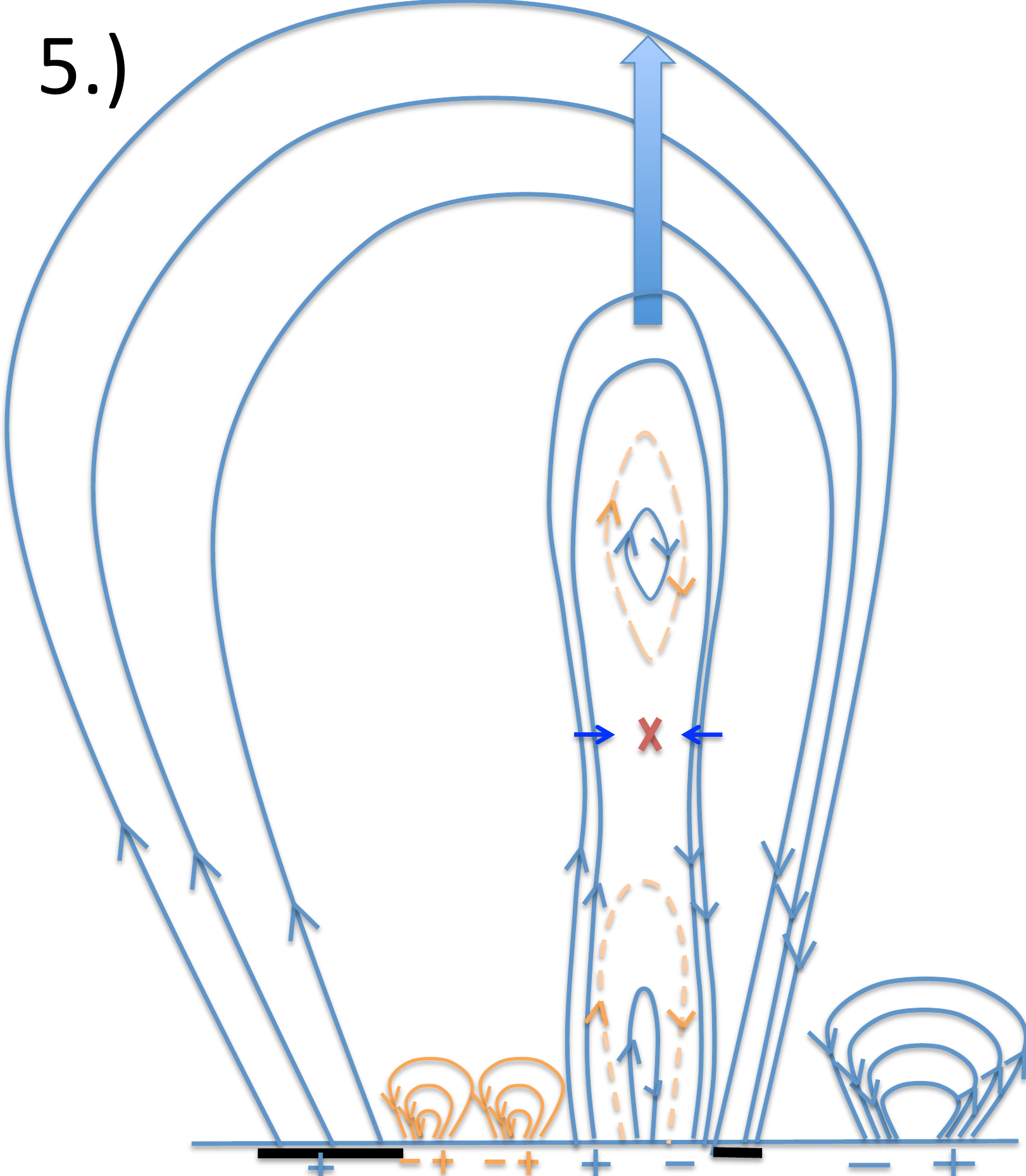}
\includegraphics[scale=0.26, trim=190 30 240 30, clip]{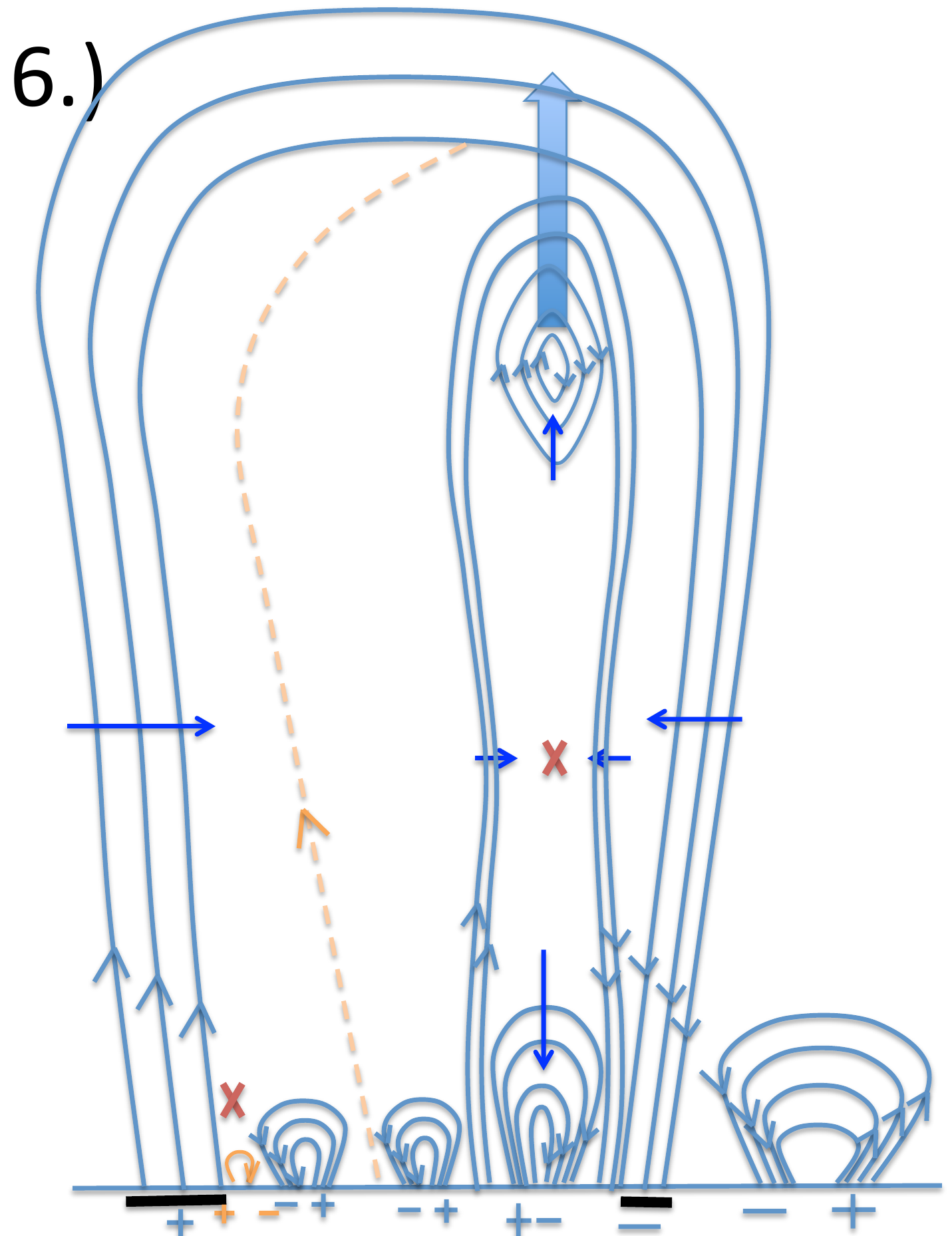}}
\centerline{\hspace*{0.015\textwidth} 
\includegraphics[scale=0.29]{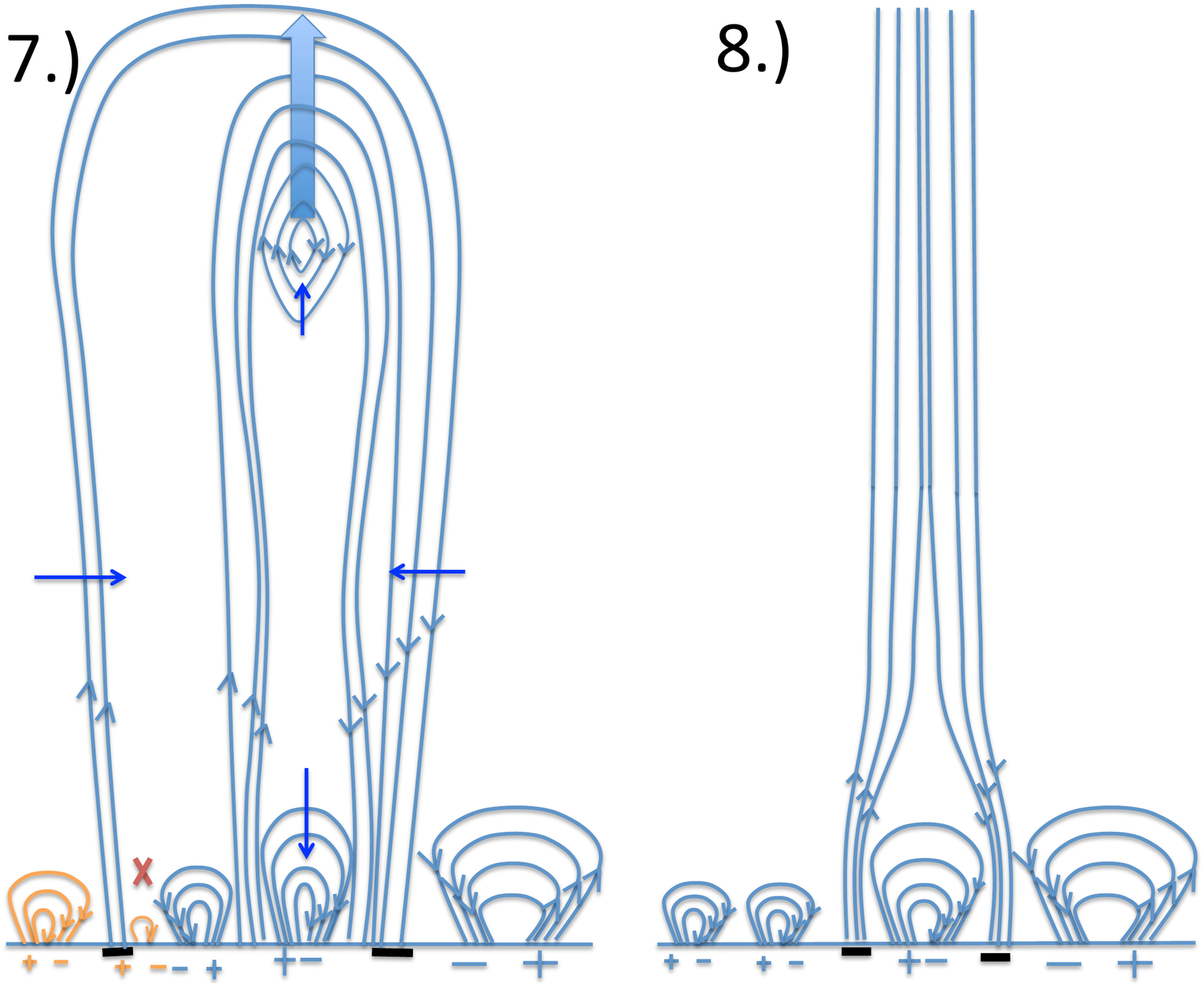}} %(top, left , bottom) trim=120 50 100 40, clip angle=-90,
%\vspace{-0.01\textwidth} % Shift back to the panel bottom 
\caption{Single dimming expansion and shrinkage explained with interchange reconnection. The blue field lines show the existing magnetic field lines, the orange lines show the newly displaced field lines after reconnection (shown with red X). The polarity of the magnetic field line footpoints is indicated with + and -- signs at the bottom of each panel. The dark blue arrows indicate the movement of the magnetic field lines. The thick black lines under the loops show the location and extent of the dimming region. For detailed description see the text.}
\label{sketch}
\end{figure} 

Figure \ref{sketch} shows a sketch of how a single dimming region could appear, expand and then shrink (as seen in our observations). Step 1 shows a simple approximation of the magnetic field configuration in a north-south cut across the AR. On the left we can see a couple of small dipole regions that represent the numerous small dipole regions in the QS (as seen south of AR 11305). We specifically chose the small scale magnetic field orientation to demonstrate the reconnection process, but we expect the small bipoles in the QS to be randomly oriented and hence the magnetic field lines will reconnect in all directions where the appropriate magnetic orientation occurs. The dipole region on the right (i.e., AR 11305) represents the loop structures connecting strong magnetic fields, with similar polarity footpoints nearby (i.e., north of AR 11305). This magnetic configuration inhibits the spreading of the CME footpoints (i.e., dimming region) to the right (i.e., north of AR 11305). Due to a nearby triggering effect (e.g., flare in a nearby AR) the loop structure shown in the middle destabilizes and starts to expand (step 2). Since it is free to expand into the QS, it gradually reconnects with the small dipole structures of opposite magnetic field direction. Through interchange reconnection the outer magnetic field lines of the AR loop structure reconnect with the outer field lines of the small bipoles--allowing the AR loop footpoints to move outwards (southward on the solar disk). As a result, oppositely oriented small structures are created at the former footpoint of the displaced AR footpoint (steps 3 and 4). As the AR loops reconnect with the small loops, the large loops in a sense ``peel off" the field lines of the small loop structure and build up a small oppositely oriented dipole on its right hand-side. The same process repeats until stability is reached temporarily (step 5). At this point the dimmings appear most disconnected and furthest spread out from the AR. The dimming location is indicated with a thick black line under the expanding loop footpoints. Our observations are consistent with the idea that the magnitude of the eruption and thus the CME mass affect the relative sizes of the consecutive dimmings--perhaps through increased energy allowing more interchange reconnection.

Starting with step 4 reconnection is shown to take place in the core of the AR loop structure as the field lines expand and extend in altitude. A plasmoid is created and the AR loops are dragged out into the heliosphere. As the plasmoid gains distance from the footpoints the magnetic curvature forces cause the loop legs to move inward towards the loop center (steps 6 and 7). This consequently leads to interchange-reconnection with small loop structures in the reverse direction. As a result the AR loop footpoints return to a pre-disturbance magnetic configuration. However, some of the field lines may be dragged out far enough that they essentially become permanently ``open", and hence, two small open magnetic field regions linger until they are dispersed (step 8). In some cases (dimming event 3, for example), the dimming forms before the previous dimming has completely receded. In Figure 5 this would correspond to a subsequent eruption at step 6 or 7. In those cases, the resulting dimming is some combination of the current and previous dimming. This is not likely to be a large effect, as the previous dimming will be recovering during the subsequent eruption and (in our examples) does not seem to dominate the signal. It is also not clear that base-difference images would help in this case, as a base-difference would not include the area around the previous dimming at all, even if some recovery had taken place.

We would like to note that this hypothesis of the creation, expansion and shrinkage of a single dimming region is a simplified, generalized description. This proposed configuration could be extended to apply to an arcade of loops in 3D. It is also possible that interchange reconnection could happen on multiple fronts in a complicated AR, creating multiple adjacent footpoints, which may be observed as a large, single dimming with mixed polarity (similar to what we observed).

In the present paper we introduce a new dimming detection tool (CODIT) that was developed from a CH detection method. Difference images, which are often used to detect dimmings, do not allow the precise determination of dimming boundaries at a given time and are dependent on the chosen pre-eruption base image (from which the consequent images are subtracted). Our method, however, directly detects dimming regions in EUV images based on their lower intensities relative to the surrounding QS. Determining the dimming boundaries at exact times means we are able to track the real evolution of the dimming boundary. Consequently we can track the evolution of the CME footpoint locations and the magnetic field distribution within the detected regions. We used CODIT to study an AR which produced six dimmings, six flares and four observed CMEs, the properties of which showed a similar trend. It appears that with each eruption the AR released more energy and mass until its last, smaller eruption. The eruptions were all linked to a recurring single dimming region. We suggest that the dimming region repeatedly expanded and shrank due to interchange reconnection displacing the footpoint locations of the CME flux rope. 

While this study found a similar trend in the flare magnitudes, dimming areas and CME masses, a large scale study is needed to truly confirm this relationship. Future work will include analysing numerous dimmings from an AIA dimming catalogue compiled by the author. In addition, dimming detections will be linked to in-situ solar wind data and geomagnetic indices to better understand the relationship between dimmings, solar wind properties and the scale of geomagnetic disturbances. We will also develop an automated version of CODIT to allow real-time dimming detection for space weather forecasting purposes.

\acknowledgments
We thank Sarah Gibson, Giuliana de Toma, Joan Burkepile, Doug Biesecker, Vic Pizzo and Howard Singer for helpful and insightful discussions. We also thank the AIA and HMI science teams for the SDO/AIA and HMI data and the STEREO/SECCHI team for the COR2 and EUVI data. This work is supported by NSF/SHINE grant AGS-0962664.

\clearpage

%% the file, uncomment the \plotone, \plottwo, and
%% If you need a layout that cannot be achieved with \plotone or
%% \plottwo, you can invoke the graphicx package directly with the
%% \includegraphics command or use \plotfiddle. For more information,
%% please see the tutorial on "Using Electronic Art with AASTeX" in the
%% documentation section at the AASTeX Web site,
%% http://www.journals.uchicago.edu/AAS/AASTeX.
%%
%% The examples below also include sample markup for submission of
%% supplemental electronic materials. As always, be sure to check
%% the instructions to authors for the journal you are submitting to
%% for specific submissions guidelines as they vary from

%% This example uses \plotone to include an EPS file scaled to
%% 80% of its natural size with \epsscale. Its caption
%% has been written to indicate that additional figure parts will be
%% available in the electronic journal.

\end{document}